\documentclass[prl,superscriptaddress,amsmath,amssymb,showpacs,noshowkeys,a4paper,preprint]{revtex4}

\usepackage{graphicx}% Include figure files
\usepackage{dcolumn}% Align table columns on decimal point
\usepackage{bm}% bold math
\usepackage[usenames]{color}
\usepackage [babel=true]{csquotes}
\usepackage{MnSymbol}
\usepackage{epstopdf}
\definecolor{mygray}{gray}{0.5}
\usepackage{enumerate}
\usepackage{chemfig}
\usepackage{ulem}

\newcommand\setpolymerdelim[2]{\def\delimleft{#1}\def\delimright{#2}}
\def\makebraces[#1,#2]#3#4#5{%
\edef\delimhalfdim{\the\dimexpr(#1+#2)/2}%
\edef\delimvshift{\the\dimexpr(#1-#2)/2}%
\chemmove{%
\node[at=(#4),yshift=(\delimvshift)]
{$\left\delimleft\vrule height\delimhalfdim depth\delimhalfdim
width0pt\right.$};%
\node[at=(#5),yshift=(\delimvshift)]
{$\left.\vrule height\delimhalfdim depth\delimhalfdim
width0pt\right\delimright_{\rlap{$\scriptstyle#3$}}$};}}
\setpolymerdelim[]

\begin{document}

\title{Walks of bubbles on a hot wire in a liquid bath. }

\author{A.~Duchesne}
%\email{alexis.duchesne@ulg.ac.be}
\affiliation{GRASP, UR-CESAM - Physics Department B5, University of Li\`ege, B-4000 Li\`ege, Belgium.}
\author{H.~Caps}
\email{herve.caps@ulg.ac.be}
\affiliation{GRASP, UR-CESAM - Physics Department B5, University of Li\`ege, B-4000 Li\`ege, Belgium.}

\begin{abstract}
When a horizontal resistive wire is heated up to the boiling point in a subcooled liquid bath, some vapor bubbles nucleate on its surface. Traditional nucleate boiling theory predicts that bubbles generated from active nucleate sites, grow up and depart from the heating surface due to buoyancy and inertia. However, we observed here a different behavior: the bubbles slide along the heated wire. In this situation, unexpected regimes are observed; from the simple sliding motion to bubble clustering. We noticed that bubbles could rapidly change their moving direction and may also interact. Finally, we propose an interpretation for both the attraction between the bubbles and the wire and for the motion of the bubbles on the wire in term of Marangoni effects. 

\end{abstract}

\maketitle

The self propulsion of a droplet actually encounters a growing interest. It has been shown that it is possible to move a droplet in a  controlled way using magnetic \cite{katsikis_2015} or electrical \cite{brandenbourger_2016} fields, vibrating bath \cite{from_bouncing_to_floating}, circular hydraulic jump \cite {Duchesne_2013} but also by changing the wetting condition \cite{style_2013, haefner_2015} or even the geometry \cite{lorenceau_2004}. An other propulsion mechanism is the Marangoni effect: it could be thermal  \cite{brochard_1989, yarin_2002, bjelobrk_2016} or chemical \cite{brochard_1989, manoj_1992, sumino_2005, oshima_2014}. But surprisingly only few papers considered self propulsion mechanisms for bubbles (this question is for instance marginally presented in \cite{kannengieser_2010}).

In the present letter, we will address the question of self propelled bubbles on a heated wire immersed in a liquid bath. The bubbles we consider are generated by nucleate boiling on the wire but, as shown, our results are more generally applicable.

\begin{figure}[t]
\begin{centering}
\includegraphics[width=0.8 \columnwidth]{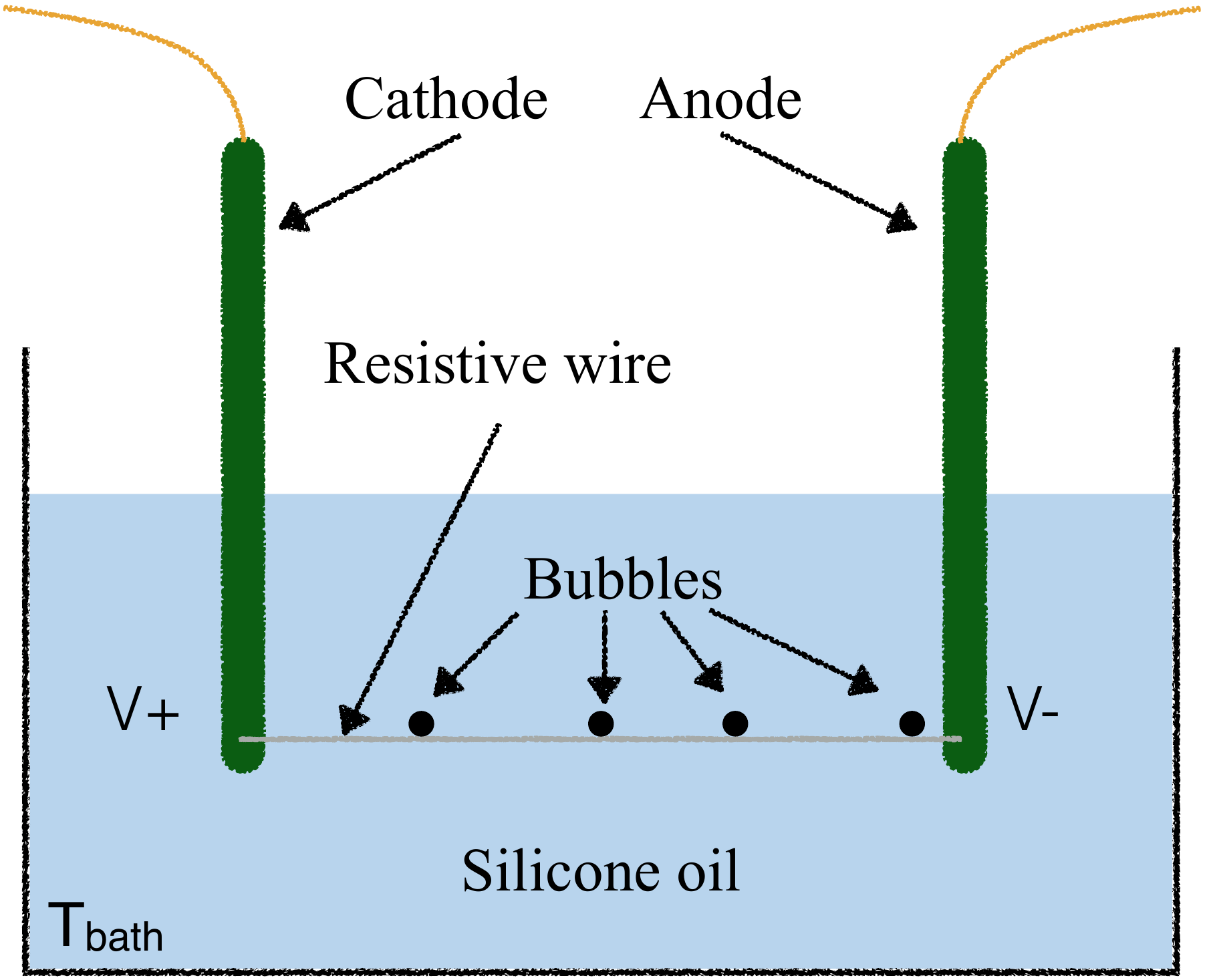}
\caption{\label{fig1} A 5 mm long resistive wire of diameter $\phi$ is fed by an electric generator through two aluminum electrodes. The wire is immersed in a silicone oil (1.5 cS) liquid bath maintained at $T_{bath}$. If a sufficient current is injected, the wire will heat thank to Joule's effect and bubbles will nucleate on its surface.}
\end{centering}
\end{figure}

A sketch of the experimental setup is depicted on Fig.\ref{fig1}. A horizontal resistive wire (constantan) immersed in a liquid bath is fed by an electric generator through two large aluminum electrodes with a negligible resistivity. The wire is thus heated up through Joule's effect with current setting up to $64$~A so that the injected power $P$ may reach $200$~W.

The wire is $5$~cm long and its diameter $\phi$ was varied from $0.1$ to $1$~mm. The surrounding liquid bath is constituted by $1.5$~L of low viscosity silicone oil ($1.5$~cS) and maintained at a constant temperature $T_{bath}$. In order to keep a constant temperature for this liquid bath it was thermalized using a water-bath. This setup allowed us to vary the bath temperature $T_{bath}$ from $5$ to $95^\circ$C. The use of silicone oil as surrounding fluid guarantees the total wetting of the liquid on the wire. 

The setup is completed by a fast camera acquiring at a rate of $1000$~fps placed in front of the wire in a horizontal plane. 

\begin{figure}[!h]
\begin{centering}
\includegraphics[width=1 \columnwidth]{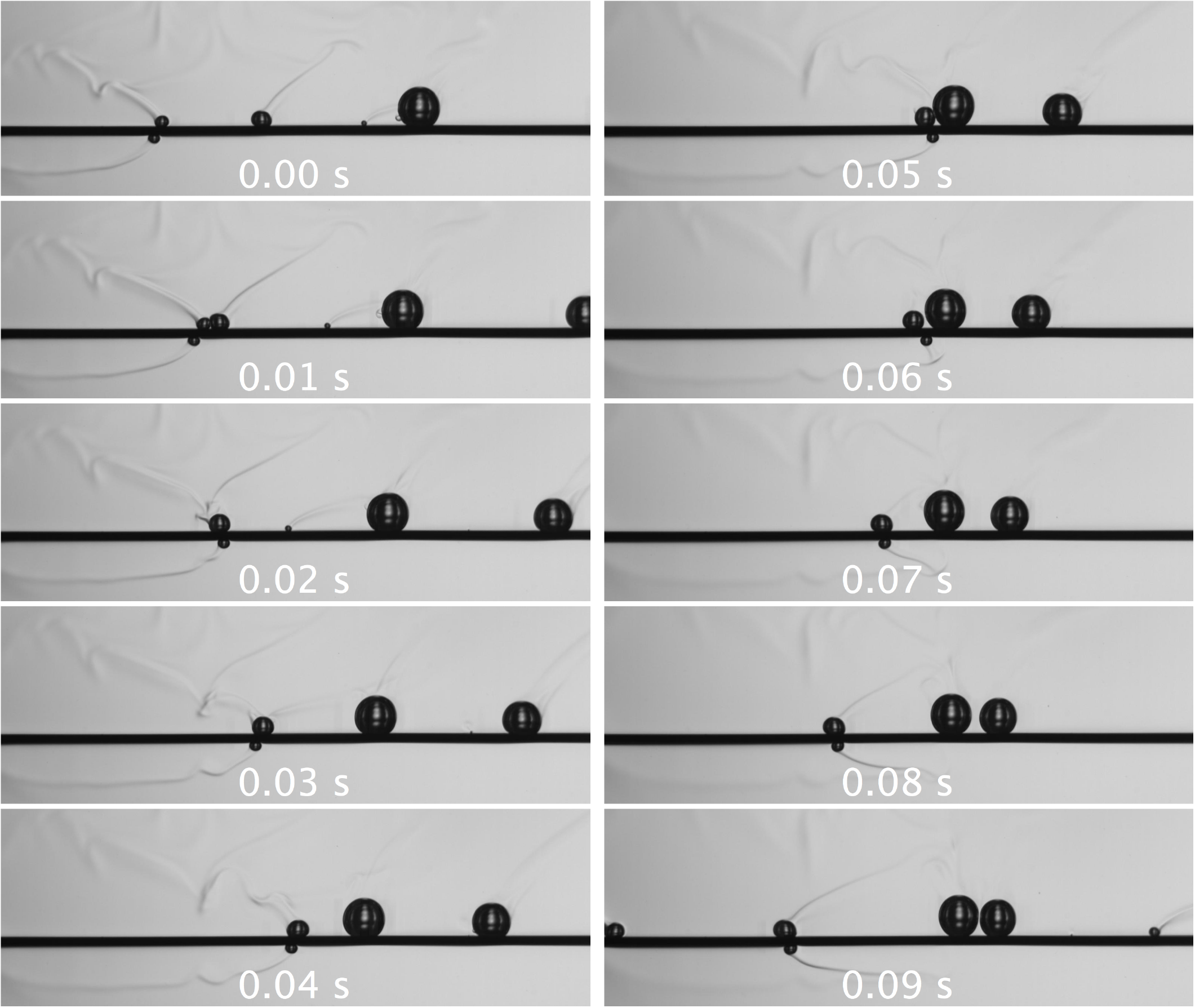}
\includegraphics[width=0.9 \columnwidth]{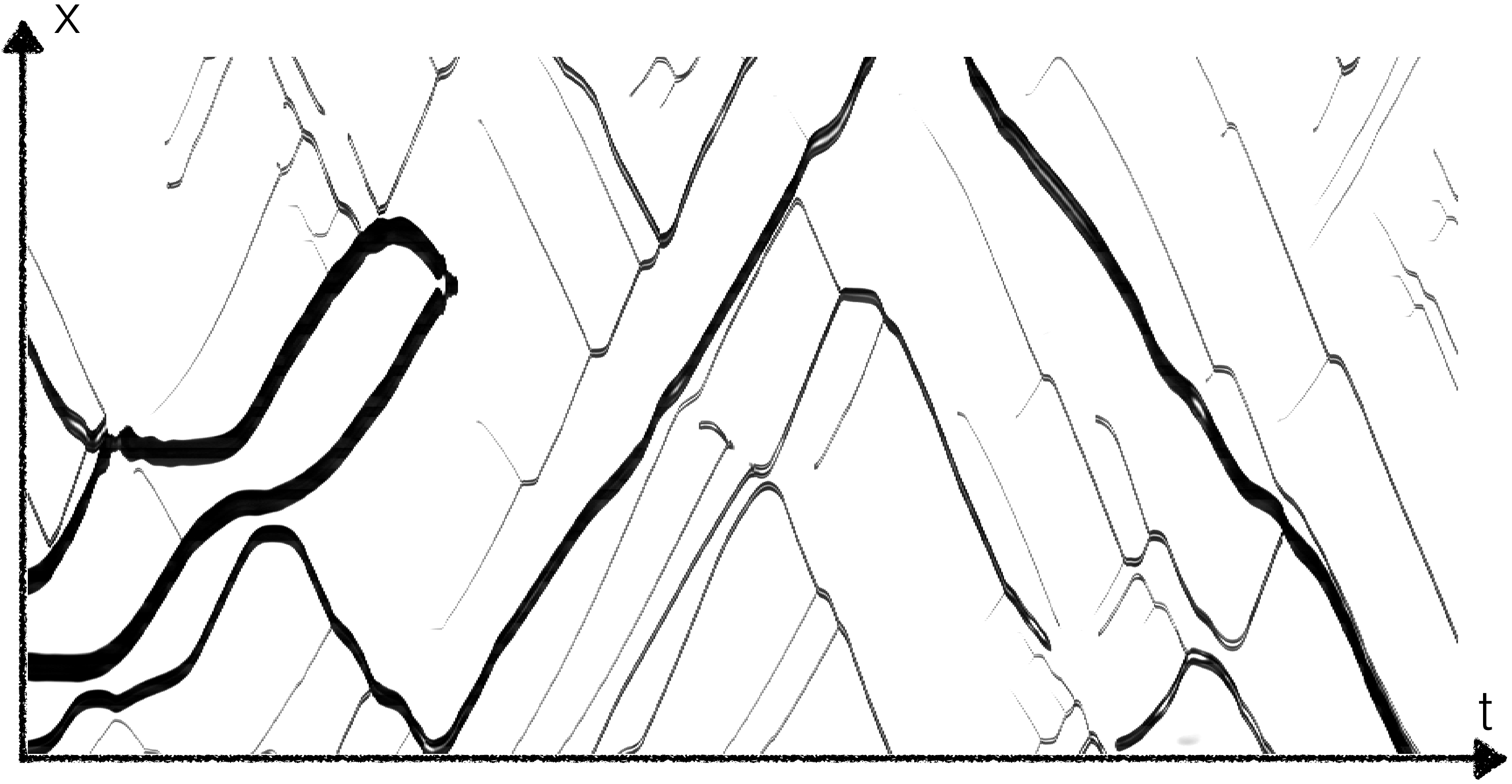}
\caption{\label{fig2} Bubbles moving on a hot wire in ``free motion'' regime ($\phi=0.2\,\mathrm{mm}$ and $P=10\, \mathrm{W}$). (a) Succession of images of bubbles moving on the wire. Images are separated by 0.01~$\mathrm{s}$. One can notice the inclined thermal plume as a signature of the bubble motion. When two bubbles comes into contact they can bounce (between 0.04~$\mathrm{s}$ and 0.07 $\mathrm{s}$) or fuse (between 0.00 $\mathrm{s}$ and 0.02 $\mathrm{s}$).  (b) Spatio-temporal diagram ($t$ in abscise and $x$ in ordinate). The pixel strip distance from the wire top surface is $50\,\mathrm{\mu m}$, its total length is $13\,\mathrm{mm}$ and the acquisition time was $0.88\, \mathrm{s}$. The thin black curves correspond to small bubbles (only the bubble top is in the pixel strip) and large black curves are signature of large bubbles. One can notice that all the bubbles have maximum velocities of the same order of magnitude.}
\end{centering}
\end{figure}

When such a metallic wire is heated up to the boiling point in a subcooled liquid bath some vapor bubbles nucleate on its surface. In the literature, it is admitted that these bubbles, generated from active nucleate sites, grow up and depart from the heating surface, experiencing buoyancy, inertia and wetting forces (e.g. \cite{hahne_1977}).  

What we observed it that beyond a threshold in injected power $P$ (corresponding to the silicone oil boiling temperature) bubbles effectively nucleate on the wire surface but instead of what is usually expected, they immediately begin to move along the wire. Typical images of this bubble motion are depicted on Fig.\ref{fig2} (a) and can be seen on videos in the supplementary material \cite{material}. One can see on the pictures a thermal plume attached to each bubble. These plumes are the signature of the liquid refractive index variation due to heating. One can also notice that bubbles are interacting: when two bubbles comes into contact they can bounce (see Fig.\ref{fig2}(a) between 0.04 s and 0.07 s) or fuse (see the Fig.\ref{fig2}(a) between 0.00 s and 0.02 s).

Such phenomena (bubble motions, bouncing and fusion) were already observed by Peng {\it et al.} \cite{peng_2011} in an experimental set up approaching the one described here (a horizontal heated platinum sub-millimeter wire immersed in water or alcohol subcooled baths). 

In the present letter, we completely renew the description of the phenomenon and its comprehension. We rationalized the experiments by considering the experimental set up as previously described. Choosing silicone oil for the liquid bath ensured a complete wetting state and so increased deeply the reproducibility of the observed phenomena. The use of constantan permitted to avoid the change of resistivity with temperature and then let us present the first phase diagram for sliding bubbles on a heated wire. This methodical analysis led to the discovering of an original collective behavior for bubbles on a thin wire. 

Theoretical aspects of the phenomena are also addressed in the present letter by proposing answers to two apparent simple but fundamental questions :\\
\indent (i) Why do the bubbles not leave the wire? \\
\indent (ii) Where the horizontal propulsion mechanism arises from?

Beyond the boiling threshold, we called the first regime we observed ``free motion'' regime, where bubbles freely circulate on the wire. It is illustrated in Fig.\ref{fig2}. As soon as the bubbles nucleate they are observed to move. Accounting for the camera frame rate, this means that the bubble grows in contact with the wire less than $0.001$~s. Direct observations also reveal that bubbles may appear and move on top or beneath the wire, even if the top position is more stable due to Archimede's force. The experiments show that, for a horizontal wire, there is no preferential sense for the bubble displacement.

We have observed a range of bubble sizes from the resolution size of the camera ($\sim 10 \, \mathrm{\mu m}$) up to millimeter length. The bubble size always increases, either by evaporation or by fusion with an other bubble. Beyond a critical radius $\mathrm{R_{max}}$ bubbles leave the wire due to buoyancy. 

The bubble velocity can reach 100 mm/s, i.e. of the order of one hundred bubble typical diameters per second. This maximal velocity is quite independent on the bubble size, as evidenced in Fig.\ref{fig2} (b) where a spatio-temporal diagram is taken at 50 $\mathrm{\mu m}$ above the wire surface. When the ``free motion'' regime is observed, the maximal velocity does not depend on the injected power $P$ (an incertitude of $\pm 10\%$ is noticed for $P\in [10; 130]\, \mathrm{W}$) and decreases by almost 50\% between a wire of $\phi=0.1 \, \mathrm{mm}$ to a wire of  $\phi=1 \, \mathrm{mm}$ for a given $P$.

\begin{figure}[!h]
\begin{centering}
(a) \includegraphics[width=0.43 \columnwidth]{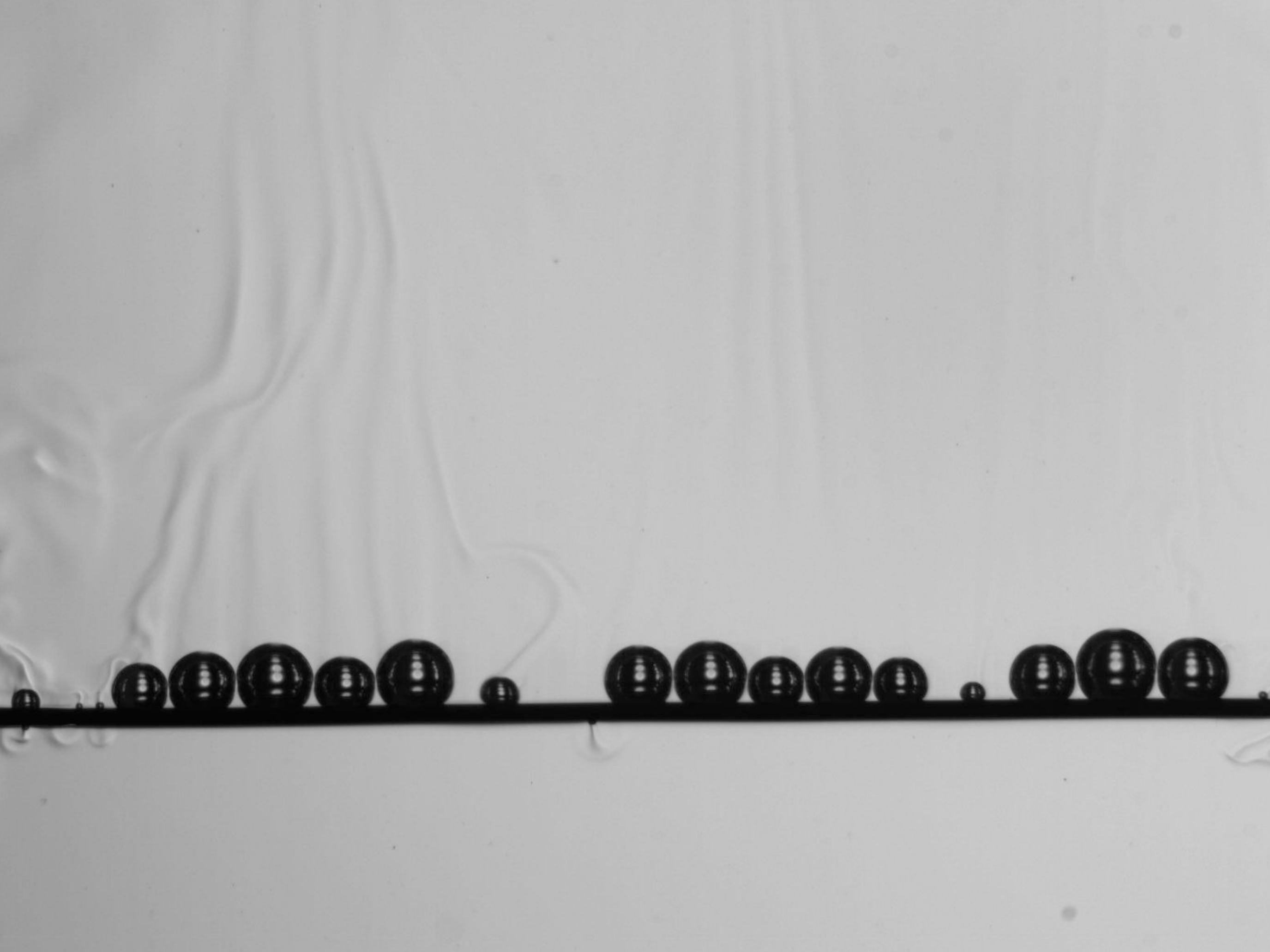}
(b) \includegraphics[width=0.43 \columnwidth]{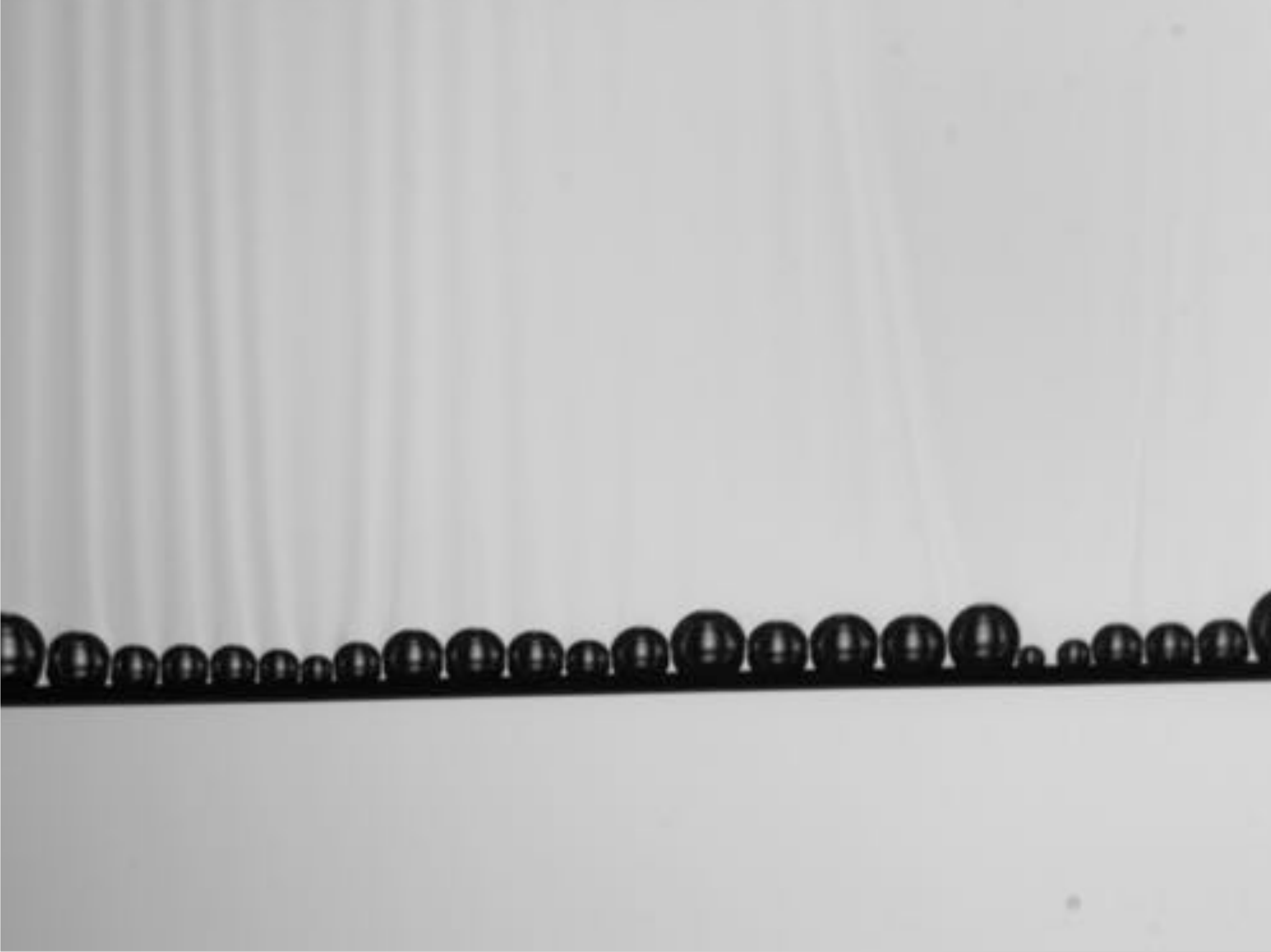}
\caption{\label{fig3} (a) ``Clusters'' regime, with $\phi=0.2$ mm and $P=17.5$ W. For this injected power $P$, the typical size for a cluster is 4-5 bubbles, small bubbles circulate between clusters. (b) The wire is covered by a single (almost) motionless cluster, with $\phi=0.2$ mm and $P=20$ W. }
\end{centering}
\end{figure}

In the middle of this ``free motion'' regime, we discovered an unexpected phenomenon: the apparition of clusters of 4 or 5 motionless bubbles. These bubbles have a quite important size ($\sim 0.5$ mm) while smaller bubbles continue to circulate between these clusters. This situation is visible on Fig.\ref{fig3}(a)  [See also Supplementary materials \cite{material}].

If we further increase the heating power $P$, clusters expand over the whole wire up to cover it as a single motionless cluster (see Fig.\ref{fig3}(b)). To be more precise, it is an almost motionless cluster: due to the heat transfer, bubbles grow up may fuse and then leave the wire, they are immediately replaced by a reorganization of bubbles on the wire. However this process is very slow in comparison to our acquisition time [See Supplementary materials \cite{material}]. If we further increase the injected heat ($\phi$ unchanged), we observe that this unique cluster dislocates into smaller clusters until this regime ends and the system comes back to the situation previously described where isolated bubbles move freely, the ``free-motion" regime. 

\begin{figure}[t]
\begin{centering}
\includegraphics[width=1.0 \columnwidth]{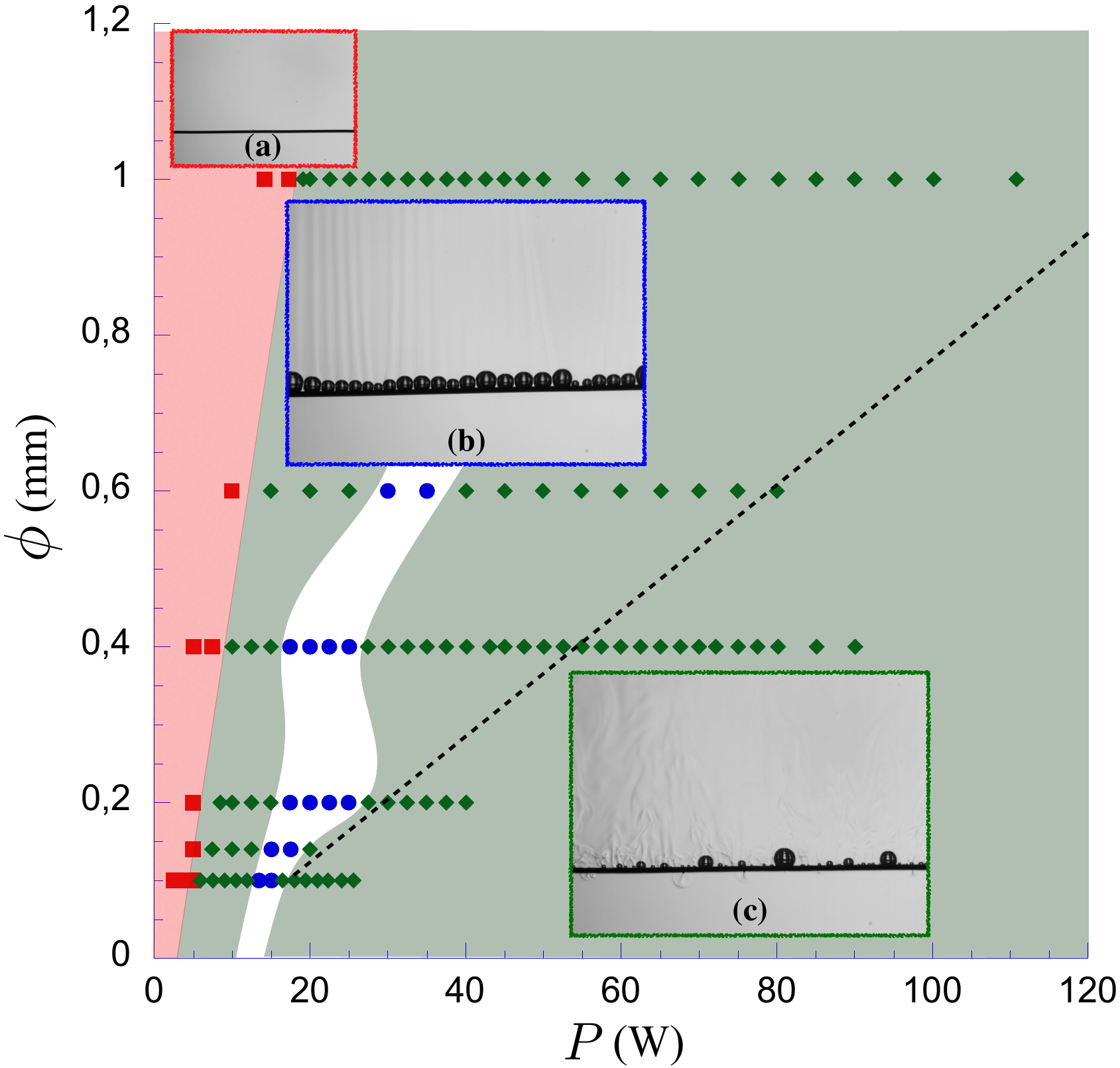}
\caption{\label{fig4} Phase diagram built with our two control parameters: the wire diameter $\phi$ and the injected power $P$. The red squares correspond to a situation without bubbles (insufficient heating powers). The green diamonds stand for the ``free motion'' regime and the blue disks for the ``clusters'' regime. The dashed line represents the limit beyond it film boiling was observed. (a) $\phi=0.2$ mm, $P < 10$ W. (b) $\phi =0.2$ mm, $P=20$ W. (c) $\phi=0.2$ mm, $P=30$ W. }
\end{centering}
\end{figure}

One can notice that beyond this ``clusters" regime the bubble density on the wire is increased and will continue to increase as long as $P$ does [see for instance, Fig.\ref{fig4}(c) but also Supplementary materials \cite{material}]. 

For the higher values of $P$ a vapor film totally isolates the wire from the liquid bath. This phenomenon is called \textit{film boiling}. Because of Rayleigh-Taylor instability, this vapor film is destabilized following a well defined wavelength \cite{lienhard_1964, son_2008}. 

A phase diagram built with our two control parameters ($\phi , P$) is presented in Fig.\ref{fig4}. The red squares correspond to heating power values insufficient to the nucleation of bubbles on the wire surface. The green diamonds stand for the ``free motion'' regime and the blue disks for the ``clusters'' regime. One can notice that for the largest wire, we do not observe ``clusters'' regime. This could be explained by the fact that for larger wire radius the bubble diameter is of the same order or below the wire diameter and so motions in depth could be observed. 

The presented measurements have been repeated in two different aquariums with different sizes. We verified that transitions between regimes do not change with the aquarium dimensions. We also checked that these transitions are not hysteretic. These informations confirmed that the physical mechanisms implied in this experiment are at the wire scale. 

Nevertheless, this is not the case for the film boiling regime. We observed a dependency on the wire dimensions, on the wire fixation system and hysteretic transitions. We did not investigate further this regime, clearly beyond the scope of this study. The dashed line on Fig.\ref{fig4} represents the limit beyond which film boiling was observed.

After this description of the different regimes, the rest of the paper will be dedicated  to the ``free motion'' regime. Particularly we would like to address the question of the interaction between the bubble and the wire. This question is not trivial: if the buoyancy leads to the bubble departure what is the force responsible for the attraction of the bubble towards the wire ? Indeed, the use of silicone oil guarantees a total wetting situation and so we can assume that a moving bubble is not in contact with the wire surface (this hypothesis is in agreement with the observations made by  Lu \& Peng \cite{lu_2006} in partial wetting situation). Therefore we can assume that the absence of contact line means no contact forces. 

We suggest to consider that the temperature differences relative to the bubble appearance are sources of Marangoni stresses responsible for both the attraction between the bubbles and the wire and the bubble motion.

\begin{figure}[!t]
\begin{centering}
\includegraphics[width=0.95 \columnwidth]{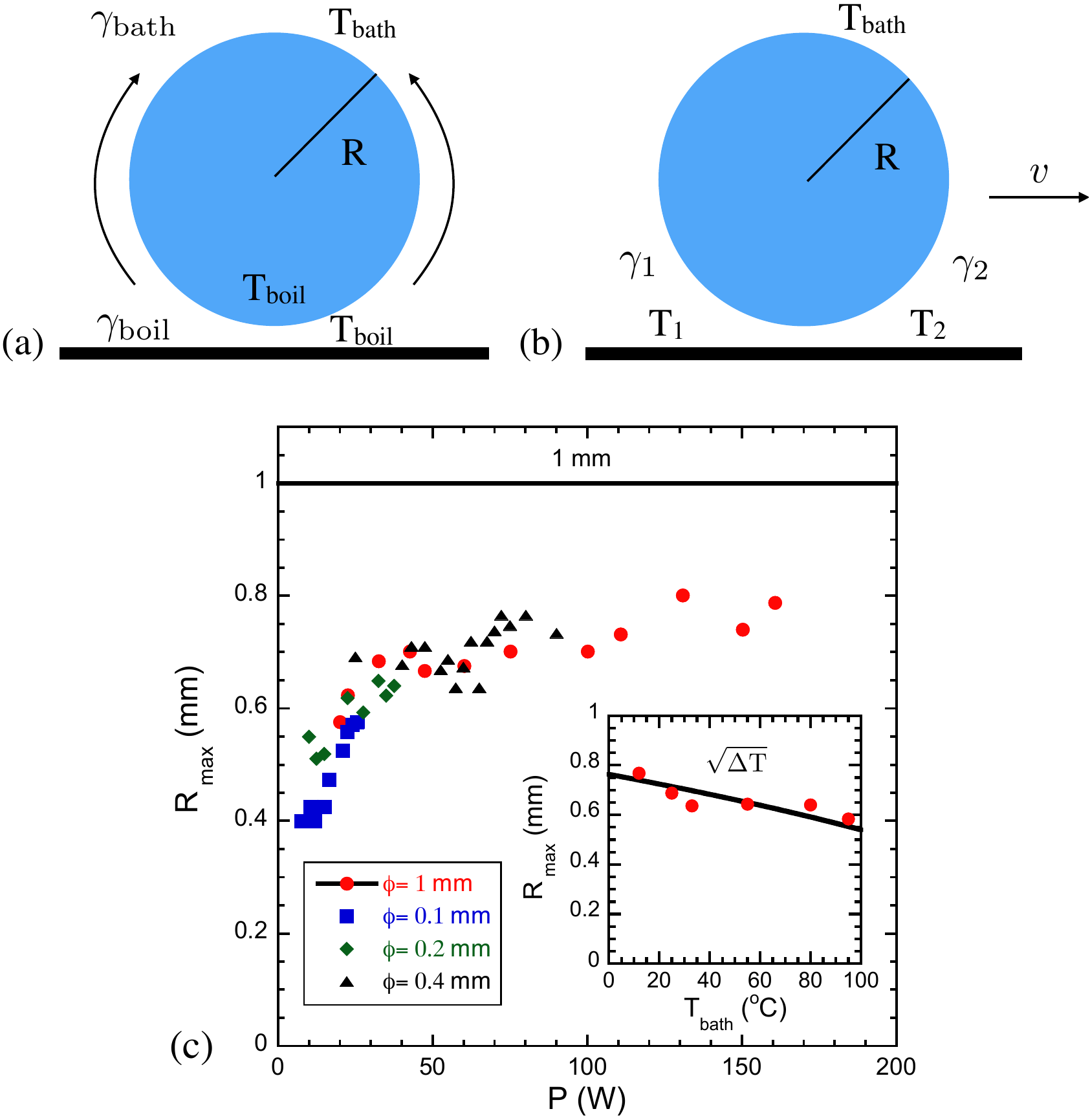}
\caption{\label{fig5} (a) Sketch showing the mechanism of attraction between the bubble and the wire: the temperature difference between the top and the bottom of the bubble induces Marangoni flows. (b) Sketch explaining the displacement of the bubble: the temperature difference between the front and the rear of the bubble induces Marangoni flows. (c) $\mathrm{R_{max}}$ versus $P$ for different $\phi$: blue squares corresponds to $\phi=0.1$ mm, green diamonds to $\phi=0.2$ mm, black triangles to $\phi=0.4$ mm and red disks to $\phi=1$ mm. The bath temperature is $\mathrm{T_{bath}}=25^\mathrm{o}\mathrm{C}$. The black continuous line is the value predicted by Eq.(\ref{e.3}). Insert: $\mathrm{R_{max}}$ versus $\mathrm{T_{bath}}$ with $\phi=0.4$ mm and $P=55$ W. The continuous black curve is a fit of the experimental data with a law in $\sqrt{\mathrm{\Delta T}}$. }
\end{centering}
\end{figure}

The bottom of the bubble is so close to the wire that the temperature at the bubble bottom should be close to the boiling temperature of the liquid, $\mathrm{T_{boil}}$, whereas the top of the bubble is further from the thermal boundary layer and so closer to the bath temperature $\mathrm{T_{bath}}$. As described in the literature \cite{young_1959}, such a temperature difference leads to surface tension differences. The latter may lead to an overpressure which might compensate buoyancy [see Fig.\ref{fig5}(a)]. 

The Marangoni's force reads: 

\begin{equation}
\label{e.1}
\bold{F_{Ma}}= \iint \limits_S \frac{\mathrm{d} \gamma}{\mathrm{d}l} \bold{\mathrm{d}S}=\iint \limits_S \gamma'\frac{\mathrm{dT}}{\mathrm{d}l} \bold{\mathrm{d}S},
\end{equation}

with $\gamma$ the surface tension and $\gamma '= \frac{\mathrm{d} \gamma}{\mathrm{dT}}$. 
This expression exhibits the need of a non null temperature coefficient of surface tension. However, an internal temperature difference cannot be achieved if the bubbles are only constituted by a pure vapor of silicon oil. Two hypothesis can then be proposed: either bubbles are constituted by dissolved gases (e.g. air) or by gases resulting from the degradation of the silicone oil with the heating. We injected air bubbles close to the wire using a syringe and observed exactly the same behaviors below and above the silicone oil boiling point [See supplementary material \cite{material}]. This complementary experience increases significantly the strength of our arguments on Marangoni's force as driving mechanism. It also emphasized the large scope of the present study, the free motion regime appearing to be a general behavior for all bubbles composed by a mixture of gases in a similar setup.  

In order to propose an order of magnitude approach, we assume that thermal gradient around the bubble is constant and reads $G=\frac{ \mathrm{T_{boil}}-\mathrm{T_{bath}}}{2R}$, where $R$ is the bubble radius. In the vertical direction, we obtain: 

\begin{equation}
\label{e.2}
F_{Ma_z}= \frac{4}{3}\pi R\gamma ' \Delta \mathrm{T} ,
\end{equation}

with  $\Delta \mathrm{T}= \mathrm{T_{boil}}-\mathrm{T_{bath}}$. Balancing this force with the buoyancy $F_{A}= \frac{4}{3}\pi \rho g R^3$ (with $\rho$ the liquid density and $g$ the acceleration due to gravity), lead us to the maximal radius before the bubble departure: 

\begin{equation}
\label{e.3}
\mathrm{R_{max}}= \sqrt{\frac{\gamma' \Delta \mathrm{T}}{\rho g}}.
\end{equation}

This law can of course be experimentally tested by measuring the radius of departure of the bubbles. Neglecting the sudden departure resulting from fusions, we take $\mathrm{T_{boil}}=200^\mathrm{o}\mathrm{C}$, $\mathrm{T_{bath}}=25^\mathrm{o}\mathrm{C}$ and $\gamma '=5.5\, 10^{-5}\, \mathrm{K^{-1}}$ \cite{hardy_1979} and we get: $\mathrm{R_{max}}\approx1$ mm.  As one can see on Fig.\ref{fig5}(c), we obtain the expected order of magnitude. Moreover all points collapse on a single master curve and a plateau appears at $\mathrm{R_{max}}\approx 0.75$ mm for the higher value of $P$. According to the proposed hypothesis, the agreement between theory and data is correct. 

We also test the dependency of $\mathrm{R_{max}}$ with ${\mathrm{\Delta T}}$ in the insert of Fig.\ref{fig5}(c). Once again we grasp the observed behavior, validating our approach this way. 

The last question we would like to address is the problem of the bubble motion on the wire. Peng \textit{et al.} proposed to interpret this phenomenon in terms of thermal Marangoni flows \cite{wang_2003_mechanism,christopher_2005, lu_2006_mechanism}: when a bubble moves, a mixing zone is generated behind it. The surrounding fluid mixes with the thermal boundary layer (typically a few hundreds of microns), resulting in a colder ($\mathrm{T_1}$) fluid behind the bubble than at its front, where the thermal boundary layer remains intact (temperature $\mathrm{T_2}$). This temperature difference induces surface tension difference and so Marangoni flows at the origin of the motion. All this information appear in the sketch presented in Fig.\ref{fig5}(b).

As far as we know,  Peng \textit{et al.} have never compared their models with direct measurements (although numerical simulations provided good agreement \cite{christopher_2006,christopher_2009}). We verified that, in our situation, these arguments were credible. We balanced the component of $\bold{F_{Ma}}$ in the wire direction with the drag friction and the creation of a dynamic meniscus (with the coefficients extracted from \cite{dubois_sliding}) and we obtained that a temperature difference of only a few degrees (typically 5~K) is enough to obtain the observed velocities. Once again the proposed mechanism seems to be a good candidate to explain our observations.  

In this letter, we evidenced different regimes for the behaviors of bubbles nucleated on the surface of a heated resistive wire immersed in a subcooled bath of silicone oil: bubbles can move along the wire and aggregate in clusters. We showed that the results obtained in the ``free motion" regime could be extended to the analogous system constituted by air bubbles on a hot thin wire in a subcooled bath. We proposed simple interpretations to the phenomena of attraction between the bubble and the wire or to the motion along the wire. The common point is thermal Marangoni flows. The comparison between these models and the experimental furnished convincing agreement, given the simplifying hypothesis we took. 

This letter opens the door to a lot of different questions touching many fields in Physics: for instance, we believed that the ``clusters'' regime can be seen as a statistical phenomenon, a simplified experience with air bubble of controlled size could be of interest. The fact that the maximal velocity appears to be independent of the bubble radius is a very puzzling result from an hydrodynamical point of view. The interactions between bubbles, bouncing or fusion, stay also an open question. Moreover, preliminary experiments showed that the bubbles interact with the thermal boundary layer, does it constitute a new system with memory effects ? To conclude, the thermal point of view of this experiment is also full of questions, among others: how the thermal transfer is influenced by this mechanism ? 

\acknowledgements{Acknowledgements. This project has been financially supported by ARC SuperCool contract of the University of Li\`ege under reference ARC 11/16-03. M. M\'elard and S. Rondia are particularly thanked for technical support.
}
\bibliography{bibliojump.bib}
\end{document}